\documentclass[conference]{IEEEtran}
\IEEEoverridecommandlockouts
\usepackage{cite}
\usepackage{amsmath,amssymb,amsfonts}
\usepackage{algorithmic}
\usepackage{graphicx}
\usepackage{textcomp}
\usepackage{xcolor}
\usepackage{multirow}
\def\BibTeX{{\rm B\kern-.05em{\sc i\kern-.025em b}\kern-.08em
    T\kern-.1667em\lower.7ex\hbox{E}\kern-.125emX}}
\begin{document}

\title{Dynamic Stress Detection: A Study of Temporal Progression Modelling of Stress in Speech\\

\thanks{Thanks to Singapore Maritime Institute (SMI)}
}

\author{\IEEEauthorblockN{Anonymous Submission}}
\author{\IEEEauthorblockN{Vishakha Lall}
\IEEEauthorblockA{\textit{Centre of Excellence in Maritime Safety} \\
\textit{Singapore Polytechnic}\\
Singapore \\
vishakha\_lall@sp.edu.sg}
\and
\IEEEauthorblockN{Yisi Liu}
\IEEEauthorblockA{\textit{Centre of Excellence in Maritime Safety} \\
\textit{Singapore Polytechnic}\\
Singapore \\
liu\_yisi@sp.edu.sg}
}

\maketitle

\begin{abstract}
Detecting psychological stress from speech is critical in high-pressure settings. While prior work has leveraged acoustic features for stress detection, most treat stress as a static label. In this work, we model stress as a temporally evolving phenomenon influenced by historical emotional state. We propose a dynamic labelling strategy that derives fine-grained stress annotations from emotional labels and introduce cross-attention-based sequential models—a Unidirectional LSTM and a Transformer Encoder—to capture temporal stress progression. Our approach achieves notable accuracy gains on MuSE (+5\%) and StressID (+18\%) over existing baselines, and generalises well to a custom real-world dataset. These results highlight the value of modelling stress as a dynamic construct in speech.
\end{abstract}


\section{Introduction}
Occupational stress significantly impacts productivity and mental well-being, particularly in high-pressure, high-stakes domains such as air traffic control and vessel traffic system operations. In such environments, effective stress monitoring could play a critical role in preventing stress-induced errors and long-term psychological strain. While biosignal-based systems (e.g., electroencephalogram, heart rate variability, skin conductance) remain the most accurate for stress detection, they typically require wearable devices, which are intrusive and impractical for prolonged use. As a result, speech-based stress detection has gained traction as a non-intrusive and scalable alternative. These systems leverage acoustic and paralinguistic features of speech \cite{stress_pitch, non_verbal_emotions_1, non_verbal_emotions_2} to infer stress-related states.

However, most existing speech-based stress detection systems treat stress as a static feature, assigning a single label to an entire speech segment. We argue that this approach oversimplifies stress, which in reality is a temporally evolving feature, shaped by prior emotional and stress states. We present a novel framework for stress detection that models the temporal progression of stress using sequential models and temporally derived labels. Our work is grounded in the hypothesis that stress is influenced not only by immediate acoustic cues but also by emotional and stress states in the recent past. This aligns with the naturalistic experience of stress, which evolves over time rather than appearing instantaneously. Our contributions are threefold: 
\begin{enumerate}
    \item Stress Progression Labelling Framework: We propose and validate a labelling strategy that infers stress from temporally annotated emotional states. This approach addresses the absence of temporally dynamic stress annotations in existing speech datasets, enabling fine-grained modelling beyond traditional static labels.
    \item Temporal Stress Classification Models: We design and implement binary classification models based on Unidirectional LSTMs and Transformer Encoder architectures, enhanced with cross-attention mechanisms to capture sequential dependencies between acoustic features and stress states.
    \item Evaluation: We validate our hypothesis on multiple datasets, including a custom real-world dataset, and demonstrate consistent improvements in detection accuracy over baselines.
\end{enumerate}

\section{Related Work}

The field of speech emotion recognition (SER) has frequently intersected with stress detection due to the close psychological correlation between emotional states and stress. For instance, \cite{muser} demonstrated that emotion recognition can act as an auxiliary task for stress detection. Similarly, the creators of the StressID dataset \cite{stressid} reported baseline results based on SER-inspired acoustic features. Motivated by these insights, we reviewed and adapted several SER modelling strategies for stress detection. Notably, \cite{cnn_stress} employed LSTM-CNN hybrids for emotion recognition in emergency call centre recordings, yielding improved performance through sequence modelling. More recent developments in sequence-based models, including LSTMs \cite{lstm_emotion, lstm_emotion_2} and large language models (LLMs) for SER \cite{llm_ser}, further underscore the potential of sequence architectures.

Sequence-based architectures have also been explored for speech-based stress detection. For instance, \cite{muser} achieved state-of-the-art performance on the MuSE dataset \cite{muse} using a BERT-based sequence model, while \cite{lstm_speech_stress} proposed an LSTM-RNN architecture combined with feedforward layers. Despite the use of sequential models, these studies rely on single static stress labels assigned to speech segments. To our knowledge, no existing approach explicitly models the temporal progression of stress using dynamically evolving labels, a gap that our work seeks to address.

We also highlight a categorical distinction between chronic stress, which has been studied in the context of depression or long-term affective states \cite{depression_speech, depression_speech_2}, and acute stress arising from immediate arousal or task-related stimuli \cite{stress_measurement}. Our work specifically targets the latter.

\section{Dataset}
To effectively model the progression of stress, datasets with long, continuous speech recordings and fine-grained temporal stress labels are required. However, most publicly available stress datasets, such as StressID \cite{stressid} and MuSE \cite{muse} provide only static stress annotations for each speech segment. To enable benchmarking and compatibility with our temporal modelling approach, we derive stress progression labels by transforming these static labels using a temporal strategy described in Section \ref{labelling_strategy}.

To generate temporally evolving stress labels for training, we leverage publicly available emotion-labelled datasets that offer utterance-level (5 sec) emotion annotations, including CREMA-D \cite{crema_d}, RAVDESS \cite{ravdess}, and SAVEE \cite{savee}. 

To evaluate the generalisation capability of our approach in real-world settings, we collected a custom dataset at the \textless name of lab\textgreater. The dataset comprises speech recordings from 10 anonymised maritime professionals, each contributing two 45-minute sessions during simulated training injected with stress-inducing scenarios (e.g., simulated collisions, engine failures, and adverse weather conditions). Speech data was paired with temporally aligned EEG-based stress measurements from 0-7, which serves as ground truth, obtained using calibrated 14 channel EEG headsets. Calibration involved baseline (relaxed) and stress-induced tasks to ensure the reliability of EEG-derived stress levels. This dataset provides a rich and realistic depiction of stress progression over time in dynamic environments and is used exclusively for testing.

The decision to include a dataset in training or testing is determined by the temporal resolution of its stress or auxiliary emotion labels. Datasets that provide frequent, continuous annotations compatible with our windowed segmentation strategy are used for training. In contrast, datasets with only coarse or static annotations are used exclusively for testing. Among the evaluated datasets, MuSE is unique in that it contain both stress and emotion labels, albeit sampled at different rates. This dual annotation enables us to additionally validate the proposed relabelling strategy.

Table \ref{datasets} summarises the datasets used, along with their labels, characteristics, and their role in our experiments.

\begin{table}[]
\centering
\resizebox{0.49\textwidth}{!}{
\begin{tabular}{lllllll}
\textbf{Dataset} & \textbf{\begin{tabular}[c]{@{}l@{}}Stress \\ Labels\end{tabular}} & \textbf{Emotion Labels} & \textbf{\begin{tabular}[c]{@{}l@{}}Audio Clip\\ Length\end{tabular}} & \textbf{\begin{tabular}[c]{@{}l@{}}Temporal\\ Labelling\end{tabular}} & \textbf{Training} & \textbf{Testing} \\ \hline
CREMA-D \cite{crema_d} & $\times$ & \begin{tabular}[c]{@{}l@{}}Categorical with \\ intensity\end{tabular} & 5 sec & \checkmark  & \checkmark & $\times$ \\
RAVDESS \cite{ravdess} & $\times$ & \begin{tabular}[c]{@{}l@{}}Categorical with \\ intensity\end{tabular} & 5 sec & \checkmark & \checkmark & $\times$ \\
SAVEE \cite{savee} & $\times$ & Categorical & 5 sec & \checkmark & \checkmark & $\times$ \\
MuSE \cite{muse} & Binary & \begin{tabular}[c]{@{}l@{}}Valence and \\ Arousal\end{tabular} & 45 min & Validation & \checkmark & \checkmark \\
StressID \cite{stressid} & \begin{tabular}[c]{@{}l@{}}10 ordinal \\ levels\end{tabular} & \begin{tabular}[c]{@{}l@{}}Valence and \\ Arousal\end{tabular} & \begin{tabular}[c]{@{}l@{}}  1 min \\ to 5 min\end{tabular} & $\times$ & $\times$ & \checkmark \\
\begin{tabular}[c]{@{}l@{}}Custom\\ Dataset\end{tabular} & \begin{tabular}[c]{@{}l@{}}8 ordinal \\ levels\end{tabular} & $\times$ & 45 min & $\times$ & $\times$ & \checkmark 
\end{tabular}}                     

\caption{Summary of datasets}
\label{datasets}
\end{table}

\section{Proposed Method}

\subsection{Quantifying Emotions and Stress Labels}

Emotion annotations in the datasets are provided either as categorical labels (e.g., happy, angry, neutral) or using the Valence-Arousal–Dominance (VAD) framework \cite{vad}, which represents emotions along three continuous or binary dimensions: Valence (positive vs. negative affect), Arousal (low vs. high activation), Dominance (submissive vs. controlling). For this study, we use the binary VAD encoding of emotions, where each dimension is discretised into $0$ (low) or $1$ (high). Table \ref{vad_emotions} presents the binary VAD representations for the emotion categories across the datasets. Stress labels, where available, are either binary or ordinal. In the latter case, we apply thresholding to get binary labels. We apply the binary VAD encodings \cite{stressvad} to stress labels when $True$. Our approach is consistent with recent research that uses dimensional emotion representations (particularly arousal) as proxies for stress \cite{stress_voice_thesis, stress_nlp, stressid}. By leveraging VAD as an intermediate representation, we bridge emotion and stress labels. 

\begin{table}[]
\centering
\begin{tabular}{llll}
\textbf{Emotion} & \textbf{Valence} & \textbf{Arousal} & \textbf{Dominance} \\ \hline
\textit{Happiness} & 1 & 1 & 1 \\
\textit{Sadness} & 0 & 0 & 0 \\
\textit{Anger} & 0 & 1 & 1 \\
\textit{Fear} & 0 & 1 & 0 \\
\textit{Disgust} & 0 & 1 & 1 \\
\textit{Stress} & 0 & 1 & 0
\end{tabular}
\caption{Binary encoded VAD \cite{vad, stressvad} representation}
\label{vad_emotions}
\end{table}

\subsection{Data Preprocessing}

\begin{figure}[t]
  \centering
  \includegraphics[width=0.8\linewidth]{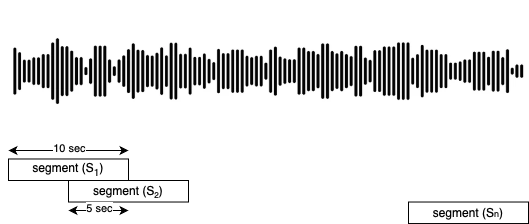}
  \caption{Temporal segmentation of long audio sequences}
  \label{segmentation}
\end{figure}

To enable dynamic stress modelling, continuous speech sequences are divided into fixed-length overlapping windows of 10 seconds, with a 5-second overlap, while preserving the corresponding labels, following prior work \cite{cnn_stress}. This windowing strategy helps retain temporal context while increasing the number of training samples. Figure~\ref{segmentation} illustrates the segmentation process.

While datasets with longer recordings(StressID, MuSE, and the Custom Dataset) readily support temporal segmentation, datasets like CREMA-D, RAVDESS, and SAVEE pose challenges due to their brevity. To address this, we implement a data augmentation strategy based on sample concatenation to simulate temporal progression. Specifically, we concatenate utterances from the same speaker and with identical linguistic content, but expressed in different emotional states. For example, utterance A spoken in a happy tone is concatenated with utterance A spoken in a disgusted tone. The resulting segment is assigned the label corresponding to the final emotional state, relying on the overlapping window strategy to capture transitions across emotional boundaries. This approach ensures the preservation of speaker identity and lexical consistency, allowing the model to focus on paralinguistic cues, such as prosody and voice quality, rather than textual information.

Effective stress detection depends on robust feature representations of speech. We explore both handcrafted and pretrained feature extraction methods. Mel Frequency Cepstral Coefficients (MFCCs) and their temporal derivatives have long been established as reliable features in speech processing, including stress detection tasks \cite{mfcc, mfcc_stress, nn_stress, cnn_speech_stress_1, cnn_speech_stress_2, stressid, lstm_speech_stress}. In addition, we experiment with pretrained deep representations: Wav2Vec 2.0 \cite{wav2vec}, used as a baseline in \cite{stressid}, and HuBERT \cite{hubert, hubert_feature_extraction}, which has shown promise in recent speech emotion recognition tasks. In our experiments, we systematically compare the performance of models trained using MFCCs, Wav2Vec 2.0, and HuBERT features to assess the impact of feature representation on dynamic stress prediction.

\subsection{Labelling Strategy}
\label{labelling_strategy}
To enable temporal stress modelling, we require a stress label for each speech segment. While our preprocessed training datasets contain emotion labels at a temporal resolution of 10 seconds, they do not include corresponding temporal stress labels. We therefore derive proxy stress labels from emotion sequences using a distance-based relabelling strategy. For each segment window $W_t$ at time $t$, the corresponding emotion label is encoded based on its binary VAD encoding $E_t$ from Table \ref{vad_emotions}. Additionally, the Hamming distance between the current VAD encoding $E_t$ and the canonical stress encoding $S$ is computed using Eq. \ref{distance_calculation}.

\begin{align}
    D_t = HammingDistance(S,E_t)
\label{distance_calculation}
\end{align}
where $HammingDistance(x,y)$ is the count of positions where $x_i \neq y_i$. To assign temporal weights to a previous segment window at time $t'$, we introduce a decaying weight.

\begin{align}
\delta_{t'} = e ^ {-\lambda(t-t')}
\label{decaying_weight}
\end{align}
where, $\lambda$ is the decay factor. Using these, a weighted distance $\theta_{t'}$ is calculated for each previous segment window. The total weighted distance at $t$ is calculated in Eq. \ref{total}.

\begin{align}
\theta_{total} = \sum_{W_{t'}=\{t, t-1...,t-n\}} \theta_{t'} = \sum_{W_{t'}=\{t, t-1...,t-n\}} \delta_{t'} \times D_{t'}
\label{total}
\end{align}
where, $n$ is the number of previous segment windows considered. The label for stress is computed in Eq. \ref{binary_stress}.

\begin{align}
\text{Label}_t =
\begin{cases}
S, & \text{if } \theta_{total} \leq T_{\text{stress}} \\
E_t, & \text{if } \theta_{total} > T_{\text{stress}}
\end{cases}
\label{binary_stress}
\end{align}

The threshold $T_{\text{stress}}$ is empirically determined using the range of possible values of $\theta_{total}$, which depends on the window size. As indicated in Table \ref{vad_emotions}, the Hamming distance $D_{t'}$ between the stress encoding $S$ and the VAD encoding of emotion $E_{t'}$ can range between $D_{t'}^{min} = 0$ and $D_{t'}^{max} = 2$. Therefore, the total weighted distance $\theta_{total}$, which is the sum of the weighted distances across the previous $n$ windows, will vary within a range influenced by $n$ and $D_{t'}^{max}$. 

Label assignment is applied across the full sequence. For initial windows with fewer than 
$n$ prior segments, we use all available past segments.

We conduct experiments by varying both $n$ and $\lambda$ to study their influence on downstream stress detection accuracy and temporal responsiveness.

The labelling strategy is used only during training to encode stress progression. At inference time, the model predicts stress labels directly from speech segments without computing temporal distances, relying on the temporal patterns it has learned during training.

\subsection{Models}
We experiment with two sequence-based architectures: a Unidirectional LSTM model and a Transformer encoder. Both architectures are enhanced with a cross-attention mechanism to learn dependencies between two sequences during training: the primary sequence, $n$  speech segments (including the current segment), and the context sequence, containing the corresponding $n-1$ stress labels (obtained via our temporal relabelling strategy). The cross-attention allows the model to condition current stress predictions not just on the speech features, but also on the stress observed in previous segments, capturing interdependencies between acoustic progression and stress evolution.

The Unidirectional Long Short-Term Memory (LSTM) network is a natural choice for modelling sequential dependencies in speech. Drawing inspiration from \cite{lstm_speech}, we design an architecture consisting of two parallel LSTM layers with 128 hidden units each—one processing the speech sequence and the other the stress label sequence. Outputs from both LSTMs are passed through a cross-attention mechanism, allowing the model to capture inter-sequence dependencies. The attention-infused representations are then concatenated and fed into a fully connected multi-label classification layer. Dropout is applied after the LSTM layers to reduce overfitting.

To further capture long-range dependencies and richer contextual interactions, we implement a Transformer Encoder architecture inspired by \cite{transformer}. Speech features are passed through a transformer encoder, while the stress context is encoded using a pretrained BERT-based encoder \cite{hubert}. A cross-attention block follows, allowing the model to attend to stress cues conditioned on the current and prior speech. The resulting representations are then passed through a linear multi-label classification layer to produce the VAD prediction.

\begin{figure}[t]
  \centering
  \includegraphics[width=\linewidth]{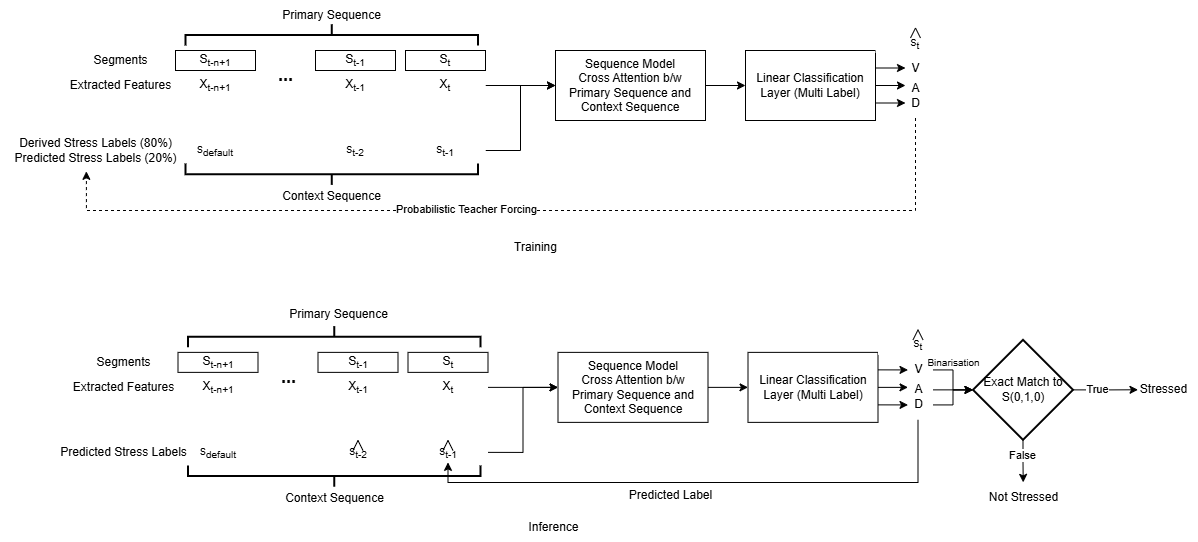}
  \caption{Model during training and inference}
  \label{model}
\end{figure}

As depicted in Figure, \ref{model}, the input to the model consists of two aligned sequences: A primary sequence of speech features $X=\{x_{t-n+1},...x_t\}$ where $x_t\in \mathbb{R}^{n\times d}$ where each $x_i$ is a $d$-dimensional feature vector (e.g., MFCC $d=40$, HuBERT/Wav2Vec $d=1024$) and a context sequence $S=\{s_{default}, s_{t-n+1},...s_{t-1}\}$ where $s_t \in \mathbb{R}^{3}$ and $s_{default} = (0,0,0)$ is added to align the sequences. The model predicts $\hat{s_t} \in \mathbb{R}^{3}$, representing the VAD stress encoding for the current segment $t$. Each component of $\hat{s_t}$ is passed through a sigmoid activation to yield a probability. During inference, these probabilities are binarised using a threshold of $0.5$ to obtain a final predicted VAD label. The stress class is determined by an exact match against the stress encoding $S=(0,1,0)$. 

The model is trained using three binary cross-entropy (BCE) losses, one for each VAD dimension, computed between the predicted $\hat{s_t}$ and ground truth $s_t$. The total loss is the average BCE across the three dimensions. Training is performed using the Adam optimiser with a learning rate scheduler and early stopping based on validation loss. Input sequences are processed in mini-batches using a sliding window over the speech and stress data, allowing the model to learn temporal dependencies and stress progression patterns.

To bridge the gap between training and inference, we incorporate probabilistic teacher forcing with a probability $p=0.8$. At each training step, the model is provided with the ground truth stress context labels $S=\{s_{t-n+1},...s_{t-1}\}$ with 80\% probability, and with 20\% probability, it uses its past prediction $S=\{\hat{s_{t-n+1}},...\hat{s_{t-1}}\}$. This scheduled sampling strategy helps the model adapt to inference-time conditions where ground truth labels are unavailable. 

The Unidirectional LSTM-based model was trained for 20 epochs, with 1000 iterations per epoch, a batch size of 16, and an initial learning rate of 0.001. The Transformer Encoder model was trained for 50 epochs, with 1000 iterations per epoch and a batch size of 16. The training was conducted on an NVIDIA GeForce RTX 4070 GPU, requiring approximately 8 hours for the Unidirectional LSTM model and 10 hours for the Transformer Encoder model. Validation metrics were monitored after each epoch to track performance.

\subsection{Evaluation Methodology}
For datasets such as MuSE and StressID, each long-form speech recording or dialogue segment is paired with a single ground-truth stress label. However, our model outputs stress predictions at 10-second intervals. To align these finer-resolution predictions with the coarse ground truth, we apply the following aggregation strategy. Each sequence is segmented into overlapping 10-second windows. The model's final stress output is used to generate each window's binary stress output label. To obtain a final prediction for the full sequence, we apply majority voting over the binary predictions of all segment windows in that sequence. This aggregated prediction is compared with the single ground-truth stress label for the full sequence to compute accuracy and F1-score.

The custom dataset is annotated with stress labels at every 10-second segment, synchronised with EEG-derived ground truth measurements. Therefore, evaluation is done at the segment level to compute accuracy and F1-score. This fine-grained evaluation reflects the model's capacity to track dynamic changes in stress within real-time operational scenarios.

\section{Results}
\subsection{Labelling Strategy Accuracy}

Table~\ref{labelling_strategy_results} reports the stress labelling accuracy obtained by varying the number of past windows $n$ and the decay factor $\lambda$. The MuSE dataset uniquely provides both discrete stress and emotion labels, making it suitable for validating our distance-based stress labelling methodology. Since emotion annotations are available at a higher temporal resolution than stress annotations, we validate our generated stress labels on speech segments closest in time (upto $n$ segments away) to the stress labels. Results show that incorporating temporal emotion context significantly improves stress label approximation. Specifically, increasing $n$ (number of past windows) and applying a moderate decay $\lambda=0.8$. yields the highest accuracy, suggesting that stress is influenced not only by the most recent emotional state but also by accumulated emotional context over time. Interestingly, accuracy consistently drops at $n=5$ across all values of $\lambda$, suggesting that emotional context beyond $40$ seconds provides diminishing or even detrimental influence on current stress estimation. These findings validate the utility of temporally evolving emotion labels as a proxy for stress progression and support our use of the temporal labelling strategy.

\begin{table}[]
\centering 
\begin{tabular}{lllll}
\textbf{\begin{tabular}[c]{@{}l@{}}$\lambda$\\ $n$\end{tabular}} & \textbf{0.01} & \textbf{0.1} & \textbf{0.8} & \textbf{1} \\ \hline
\textbf{0} & 54.3\% & 54.3\% & 54.3\% & 54.3\% \\
\textbf{1} & 57.6\% & 63\% & 78.9\% & 71.4\% \\
\textbf{2} & 58.1\% & 63.4\% & 84\% & 72.1\% \\
\textbf{3} & 58.1\% & 63.7\% & \textbf{91.2\%} & 76.7\% \\
\textbf{4} & 58.1\% & 63.7\% & \textbf{91.4\%} & 76.5\% \\
\textbf{5} & 58.1\% & 63.5\% & 90.8\% & 74.3\%
\end{tabular}
\caption{Evaluating labelling strategies by varying $\lambda$ and $n$ on MuSE}
\label{labelling_strategy_results}
\end{table}

\subsection{Stress Detection Model Performance}
Table~\ref{results} presents a comprehensive comparison of contemporary baseline models on MuSE and StressID datasets, alongside our proposed dynamic temporal models, as well as generalisation results on the Custom Dataset. The results demonstrate that our models, which incorporate temporal stress progression, consistently outperform baseline approaches across all datasets. Specifically, the Transformer Encoder architecture with HuBERT feature extraction achieves the best results across all datasets.
A notable finding is the variation in the optimal number of past windows ($n$) for different datasets. While the best performance on MuSE and the Custom Dataset is achieved with $n=4$ (40 second historical context), StressID performs best with $n=3$ (30 second historical context). This variation reflects inherent differences in the nature of the datasets: StressID includes short, stress-inducing tasks such as counting, Stroop tests, and arithmetic, whereas MuSE involves interview-style interactions, and the Custom Dataset contains high-cognitive-load scenarios like simulated emergencies. These results highlight that while temporal dependencies are critical for stress detection, the optimal extent of past context varies with the task type and recording scenario. This emphasises the need for dataset-specific tuning of temporal parameters in real-world applications of stress progression modelling.

\begin{table}[]
\centering 
\begin{tabular}{lll}
\textbf{Model}                                                                                        & \textbf{Dataset}                & \textbf{Performance}                                     \\ \hline
\begin{tabular}[c]{@{}l@{}}MLP + Opensmile\\ \cite{muse}\end{tabular}                                                                                       & \multirow{5}{*}{MuSE}           & \begin{tabular}[c]{@{}l@{}}A=0.67\\ F1=0.69\end{tabular} \\ \cline{1-1} \cline{3-3} 
\begin{tabular}[c]{@{}l@{}}MLP + LIWC\\ \cite{muse}\end{tabular}                                                                                            &                                 & \begin{tabular}[c]{@{}l@{}}A=0.60\\ F1=0.67\end{tabular} \\ \cline{1-1} \cline{3-3} 
\begin{tabular}[c]{@{}l@{}}MUSER Acoustic\\ Encoder \cite{muser}\end{tabular}                                       &                                 & \begin{tabular}[c]{@{}l@{}}A=0.79\\ F1=0.80\end{tabular} \\ \cline{1-1} \cline{3-3} 
\begin{tabular}[c]{@{}l@{}}Unidirectional LSTM\\ (n=4, Wav2Vec 2.0\\ feature extraction)\end{tabular} &                                 & \begin{tabular}[c]{@{}l@{}}A=0.81\\ F1=0.80\end{tabular} \\ \cline{1-1} \cline{3-3} 
\begin{tabular}[c]{@{}l@{}}Transformer Encoder \\ (n=4, HuBERT\\ feature extraction)\end{tabular}     &                                 & \begin{tabular}[c]{@{}l@{}}A=0.83\\ F1=0.81\end{tabular} \\ \hline
\begin{tabular}[c]{@{}l@{}}Audio-HC + kNN\\ \cite{stressid}\end{tabular}                                                                                      & \multirow{5}{*}{StressID}       & \begin{tabular}[c]{@{}l@{}}A=0.6\\ F1=0.67\end{tabular}  \\ \cline{1-1} \cline{3-3} 
\begin{tabular}[c]{@{}l@{}}Audio-DNN + SVM\\ \cite{stressid}\end{tabular}                                                                                        &                                 & \begin{tabular}[c]{@{}l@{}}A=0.54\\ F1=0.61\end{tabular} \\ \cline{1-1} \cline{3-3} 
\begin{tabular}[c]{@{}l@{}}Wav2Vec 2.0 Classifier\\ \cite{stressid}\end{tabular}                                                                                     &                                 & \begin{tabular}[c]{@{}l@{}}A=0.66\\ F1=0.7\end{tabular}  \\ \cline{1-1} \cline{3-3} 
\begin{tabular}[c]{@{}l@{}}Unidirectional LSTM\\ (n=3, Wav2Vec 2.0\\ feature extraction)\end{tabular} &                                 & \begin{tabular}[c]{@{}l@{}}A=0.75\\ F1=0.79\end{tabular} \\ \cline{1-1} \cline{3-3} 
\begin{tabular}[c]{@{}l@{}}Transformer Encoder \\ (n=3, HuBERT\\ feature extraction)\end{tabular}     &                                 & \begin{tabular}[c]{@{}l@{}}A=0.78\\ F1=0.80\end{tabular} \\ \hline
\begin{tabular}[c]{@{}l@{}}Unidirectional LSTM\\ (n=4, Wav2Vec 2.0\\ feature extraction)\end{tabular} & \multirow{2}{*}{Custom Dataset} & \begin{tabular}[c]{@{}l@{}}A=0.80\\ F1=0.80\end{tabular} \\ \cline{1-1} \cline{3-3} 
\begin{tabular}[c]{@{}l@{}}Transformer Encoder \\ (n=4, HuBERT\\ feature extraction)\end{tabular}     &                                 & \begin{tabular}[c]{@{}l@{}}A=0.81\\ F1=0.82\end{tabular} \\ 
\end{tabular}
\caption{Comparison of Stress Detection Accuracy Across Datasets and Models (A: Accuracy, F1: F1 Score)}
\label{results}
\end{table}

\subsection{Ablation Study}
We conduct an ablation study to evaluate the impact of two key design choices on model performance, window history length $n$ and feature extraction method. The evaluation is consistent across the test splits of all datasets and follows a segment-level or sequence-level protocol, depending on the dataset’s structure. 
We vary the number of past speech and stress segments $n$ provided as temporal context. The results, summarized in Table \ref{ablation_window_length}, reveal a general trend of increasing performance with larger context windows. This supports our hypothesis that stress has temporal dependencies. However, we also observe dataset-specific trends. for example, MuSE and the Custom Dataset benefit more from longer windows (e.g., $n=4$), while StressID achieves optimal performance at $n=3$. This variation aligns with the nature of the datasets: StressID contains short-form, task-specific recordings, whereas MuSE and the Custom Dataset involve conversational or scenario-driven speech with more gradual stress evolution.
We compare three types of speech feature representations: MFCC, Wav2Vec 2.0, and HuBERT. Table \ref{ablation_feature_extraction} presents the corresponding results. We find that both Wav2Vec 2.0 and HuBERT outperform traditional MFCC features across all models, highlighting the advantage of using contextualised, self-supervised embeddings for stress detection. Interestingly, Wav2Vec 2.0 achieves the best results with the LSTM-based model, while HuBERT performs best with the Transformer Encoder architecture. This could be attributed to architectural compatibility: Wav2Vec 2.0 representations, which emphasise local context and phonetic detail, align well with the sequential nature of LSTMs. In contrast, HuBERT's hierarchical clustering and token masking better capture global structure and higher-level semantics, which synergise with the Transformer's self-attention mechanism.

\begin{table*}[]
\centering
\resizebox{0.95\textwidth}{!}{
\begin{tabular}{lllllll}
\textbf{\begin{tabular}[c]{@{}l@{}}Window History \\ Length\end{tabular}}              & \multicolumn{2}{c}{\textbf{MuSE}}                                                                                                                                                                                 & \multicolumn{2}{c}{\textbf{StressID}}                                                                                                                                                                             & \multicolumn{2}{c}{\textbf{\begin{tabular}[c]{@{}c@{}}Custom\\ Dataset\end{tabular}}}                                                                                                                             \\
\textbf{}                                                                              & \textbf{\begin{tabular}[c]{@{}l@{}}Unidirectional LSTM\\ (Wav2Vec 2.0 \\ feature extraction)\end{tabular}} & \textbf{\begin{tabular}[c]{@{}l@{}}Transformer Encoder\\ (HuBERT\\ feature extraction)\end{tabular}} & \textbf{\begin{tabular}[c]{@{}l@{}}Unidirectional LSTM\\ (Wav2Vec 2.0 \\ feature extraction)\end{tabular}} & \textbf{\begin{tabular}[c]{@{}l@{}}Transformer Encoder\\ (HuBERT\\ feature extraction)\end{tabular}} & \textbf{\begin{tabular}[c]{@{}l@{}}Unidirectional LSTM\\ (Wav2Vec 2.0 \\ feature extraction)\end{tabular}} & \textbf{\begin{tabular}[c]{@{}l@{}}Transformer Encoder\\ (HuBERT\\ feature extraction)\end{tabular}} \\ \hline
\begin{tabular}[c]{@{}l@{}}n=0\\ No temporal context\\ in stress labels\end{tabular}   & \begin{tabular}[c]{@{}l@{}}A=0.57\\ F1=0.69\end{tabular}                                                   & \begin{tabular}[c]{@{}l@{}}A=0.54\\ F1=0.55\end{tabular}                                             & \begin{tabular}[c]{@{}l@{}}A=0.46\\ F1=0.46\end{tabular}                                                   & \begin{tabular}[c]{@{}l@{}}A=0.47\\ F1=0.49\end{tabular}                                             & \begin{tabular}[c]{@{}l@{}}A=0.61\\ F1=0.61\end{tabular}                                                   & \begin{tabular}[c]{@{}l@{}}A=0.61\\ F1=0.62\end{tabular}                                             \\
\begin{tabular}[c]{@{}l@{}}n=1\\ 10 s temporal context\\ in stress labels\end{tabular} & \begin{tabular}[c]{@{}l@{}}A=0.61\\ F1=0.62\end{tabular}                                                   & \begin{tabular}[c]{@{}l@{}}A=0.61\\ F1=0.61\end{tabular}                                             & \begin{tabular}[c]{@{}l@{}}A=0.55\\ F1=0.57\end{tabular}                                                   & \begin{tabular}[c]{@{}l@{}}A=0.56\\ F1=0.57\end{tabular}                                             & \begin{tabular}[c]{@{}l@{}}A=0.63\\ F1=0.64\end{tabular}                                                   & \begin{tabular}[c]{@{}l@{}}A=0.65\\ F1=0.66\end{tabular}                                             \\
\begin{tabular}[c]{@{}l@{}}n=2\\ 20 s temporal context\\ in stress labels\end{tabular} & \begin{tabular}[c]{@{}l@{}}A=0.75\\ F1=0.76\end{tabular}                                                   & \begin{tabular}[c]{@{}l@{}}A=0.76\\ F1=0.75\end{tabular}                                             & \begin{tabular}[c]{@{}l@{}}A=0.63\\ F1=0.63\end{tabular}                                                   & \begin{tabular}[c]{@{}l@{}}A=0.65\\ F1=0.64\end{tabular}                                             & \begin{tabular}[c]{@{}l@{}}A=0.69\\ F1=0.70\end{tabular}                                                   & \begin{tabular}[c]{@{}l@{}}A=0.71\\ F1=0.70\end{tabular}                                             \\
\begin{tabular}[c]{@{}l@{}}n=3\\ 30 s temporal context\\ in stress labels\end{tabular} & \begin{tabular}[c]{@{}l@{}}A=0.80\\ F1=0.79\end{tabular}                                                   & \begin{tabular}[c]{@{}l@{}}A=0.80\\ F1=0.80\end{tabular}                                             & \textbf{\begin{tabular}[c]{@{}l@{}}A=0.75\\ F1=0.79\end{tabular}}                                          & \textbf{\begin{tabular}[c]{@{}l@{}}A=0.78\\ F1=0.80\end{tabular}}                                    & \begin{tabular}[c]{@{}l@{}}A=0.77\\ F1=0.76\end{tabular}                                                   & \begin{tabular}[c]{@{}l@{}}A=0.79\\ F1=0.80\end{tabular}                                             \\
\begin{tabular}[c]{@{}l@{}}n=4\\ 40 s temporal context\\ in stress labels\end{tabular} & \textbf{\begin{tabular}[c]{@{}l@{}}A=0.81\\ F1=0.80\end{tabular}}                                          & \textbf{\begin{tabular}[c]{@{}l@{}}A=0.83\\ F1=0.81\end{tabular}}                                    & \begin{tabular}[c]{@{}l@{}}A=0.62\\ F1=0.61\end{tabular}                                                   & \begin{tabular}[c]{@{}l@{}}A=0.64\\ F1=0.64\end{tabular}                                             & \textbf{\begin{tabular}[c]{@{}l@{}}A=0.80\\ F1=0.80\end{tabular}}                                          & \textbf{\begin{tabular}[c]{@{}l@{}}A=0.81\\ F1=0.82\end{tabular}}                                    \\
\begin{tabular}[c]{@{}l@{}}n=5\\ 50 s temporal context\\ in stress labels\end{tabular} & \begin{tabular}[c]{@{}l@{}}A=0.78\\ F1=0.79\end{tabular}                                                   & \begin{tabular}[c]{@{}l@{}}A=0.79\\ F1=0.80\end{tabular}                                             & \begin{tabular}[c]{@{}l@{}}A=0.61\\ F1=0.60\end{tabular}                                                   & \begin{tabular}[c]{@{}l@{}}A=0.63\\ F1=0.64\end{tabular}                                             & \begin{tabular}[c]{@{}l@{}}A=0.75\\ F1=0.75\end{tabular}                                                   & \begin{tabular}[c]{@{}l@{}}A=0.76\\ F1=0.75\end{tabular}                                            
\end{tabular}}
\caption{Ablation Study on Window Listory Length (A: Accuracy, F1: F1 Score)}
\label{ablation_window_length}
\end{table*}

\begin{table*}[]
\resizebox{0.95\textwidth}{!}{
\begin{tabular}{lllllll}
\textbf{Feature Extraction} & \multicolumn{2}{c}{\textbf{MuSE}}                                                                                                                           & \multicolumn{2}{c}{\textbf{StressID}}                                                                                                                       & \multicolumn{2}{c}{\textbf{\begin{tabular}[c]{@{}c@{}}Custom\\ Dataset\end{tabular}}}                                                                       \\
\textbf{}                   & \textbf{\begin{tabular}[c]{@{}l@{}}Unidirectional LSTM\\ (n=4)\end{tabular}} & \textbf{\begin{tabular}[c]{@{}l@{}}Transformer Encoder\\ (n=4)\end{tabular}} & \textbf{\begin{tabular}[c]{@{}l@{}}Unidirectional LSTM\\ (n=3)\end{tabular}} & \textbf{\begin{tabular}[c]{@{}l@{}}Transformer Encoder\\ (n=3)\end{tabular}} & \textbf{\begin{tabular}[c]{@{}l@{}}Unidirectional LSTM\\ (n=4)\end{tabular}} & \textbf{\begin{tabular}[c]{@{}l@{}}Transformer Encoder\\ (n=4)\end{tabular}} \\ \hline
MFCC                        & \begin{tabular}[c]{@{}l@{}}A=0.75\\ F1=0.75\end{tabular}                     & \begin{tabular}[c]{@{}l@{}}A=0.76\\ F1=0.75\end{tabular}                     & \begin{tabular}[c]{@{}l@{}}A=0.69\\ F1=0.70\end{tabular}                     & \begin{tabular}[c]{@{}l@{}}A=0.71\\ F1=0.71\end{tabular}                     & \begin{tabular}[c]{@{}l@{}}A=0.76\\ F1=0.76\end{tabular}                     & \begin{tabular}[c]{@{}l@{}}A=0.79\\ F1=0.78\end{tabular}                     \\
Wav2Vec 2.0                 & \textbf{\begin{tabular}[c]{@{}l@{}}A=0.81\\ F1=0.80\end{tabular}}            & \begin{tabular}[c]{@{}l@{}}A=0.80\\ F1=0.79\end{tabular}                     & \textbf{\begin{tabular}[c]{@{}l@{}}A=0.75\\ F1=0.79\end{tabular}}            & \begin{tabular}[c]{@{}l@{}}A=0.76\\ F1=0.78\end{tabular}                     & \textbf{\begin{tabular}[c]{@{}l@{}}A=0.80\\ F1=0.80\end{tabular}}            & \begin{tabular}[c]{@{}l@{}}A=0.80\\ F1=0.80\end{tabular}                     \\
HuBERT                      & \begin{tabular}[c]{@{}l@{}}A=0.79\\ F1=0.78\end{tabular}                     & \textbf{\begin{tabular}[c]{@{}l@{}}A=0.83\\ F1=0.81\end{tabular}}            & \begin{tabular}[c]{@{}l@{}}A=0.73\\ F1=0.75\end{tabular}                     & \textbf{\begin{tabular}[c]{@{}l@{}}A=0.78\\ F1=0.80\end{tabular}}            & \begin{tabular}[c]{@{}l@{}}A=0.80\\ F1=0.79\end{tabular}                     & \textbf{\begin{tabular}[c]{@{}l@{}}A=0.81\\ F1=0.82\end{tabular}}           
\end{tabular}}
\caption{Ablation Study on Feature Extraction Model (A: Accuracy, F1: F1 Score)}
\label{ablation_feature_extraction}
\end{table*}

\section{Conclusion}
This work presents a novel approach to stress detection in speech by modelling its temporal progression. By introducing a distance-based labelling strategy using VAD encodings and leveraging contextual stress history via LSTM and Transformer architectures, we demonstrate significant improvements over traditional baselines. Our results affirm that stress is a temporally evolving phenomenon, and incorporating past emotional context enhances detection accuracy.

The variability in optimal temporal window lengths across datasets highlights the need to adapt temporal modelling to task-specific and contextual factors. To further enhance and validate such models, future work should explore datasets with richer temporal annotations for stress. Additionally, extending this framework to incorporate multimodal signals—such as physiological or visual data—holds promise for building more robust and comprehensive models for real-world stress detection.
\bibliographystyle{IEEEtran}
\bibliography{bibliography}

\end{document}